# The *Fermi* Gamma-Ray Burst Monitor

Charles Meegan[1], Giselher Lichti[2], P. N. Bhat[3], Elisabetta Bissaldi[2], Michael S. Briggs[3], Valerie Connaughton[3], Roland Diehl[2], Gerald Fishman[4], Jochen Greiner[2], Andrew S. Hoover[5], Alexander J. van der Horst[6], Andreas von Kienlin[2], R. Marc Kippen[5], Chryssa Kouveliotou[4], Sheila McBreen[7], W. S. Paciesas[3], Robert Preece[3], Helmut Steinle[2], Mark S. Wallace[5], Robert B. Wilson[3], Colleen Wilson-Hodge[4].

1. Universities Space Research Association, NSSTC, 320 Sparkman Drive, Huntsville, AL 35805, USA
2. Max Planck Institute for Extraterrestrial Physics, Giessenbachstrasse Postfach 1312, Garching, 85748, Germany
3. University of Alabama in Huntsville, NSSTC, 320 Sparkman Drive, Huntsville, AL 35805, USA
4. Space Science Office, VP62, NASA/Marshall Space Flight Center, Huntsville, AL 35812, USA
5. Los Alamos National Laboratory, PO Box 1663, Los Alamos, NM 87545, USA
6. NASA Postdoctoral Program Fellow, NSSTC, 320 Sparkman Drive, Huntsville, AL 35805, USA
7. University College, Dublin, Belfield, Stillorgan Road, Dublin 4, Ireland

ABSTRACT

The Gamma-Ray Burst Monitor (GBM) will significantly augment the science return from the *Fermi* Observatory in the study of Gamma-Ray Bursts (GRBs). The primary objective of GBM is to extend the energy range over which bursts are observed downward from the energy range of the Large Area Telescope (LAT) on *Fermi* into the hard X-ray range where extensive previous data sets exist. A secondary objective is to compute burst locations on-board to allow re-orienting the spacecraft so that the LAT can observe delayed emission from bright bursts. GBM uses an array of twelve sodium iodide scintillators and two bismuth germanate scintillators to detect gamma rays from ~8 keV to ~40 MeV over the full unocculted sky. The on-board trigger threshold is ~0.7 photons cm$^{-2}$ s$^{-1}$ (50 – 300 keV, 1 s peak). GBM generates on-board triggers for ~250 GRBs per year.

## 1. INTRODUCTION

### 1.1 *Gamma-Ray Bursts*

Gamma-ray bursts (GRBs) have been studied for over forty years by a multitude of international teams, from space and with ground-based observatories. Several GRB observational facts have already been established; at the same time new important questions have arisen and guided observers in designing new instrumentation to unravel this phenomenon. We now know that GRBs are the most powerful explosions in the Universe, releasing of order $10^{51}$ ergs in tens of seconds in gamma-rays. To date distances to 150 GRB sources have been established with redshifts ranging from 0.0085 (GRB 980425; Tinney et al. 1998) to ~8 (GRB0090423; Olivares et al 2009, Thoene et al. 2009, Tanvir et al. 2009). The detectability of GRBs to such high redshifts makes them important cosmological probes (e.g., Cohen & Piron 1997, Wijers et al. 1998, Tsutsui 2009). The current prevalent model on the origin of the longer (>2 s) class of bursts (Kouveliotou et al. 1993) is believed to be a core collapse supernova of a massive, rapidly rotating star – a "collapsar" (Woosley et al. 1993), while the origin of the shorter events is commonly ascribed to neutron star - neutron star or neutron star - black hole mergers (e.g., Rosswog, Ramirez-Ruiz, & Davies 2003). A recent comprehensive review of gamma-ray bursts is presented by Vedrenne and Atteia (2009).

The bulk of GRB detections before *Fermi* have been in the ~20 to 1000 keV band and much is known about the spectral properties of GRBs in this band (Kaneko et al. 2006). Spectral data are usually fit with a smoothly broken power law (Band et al. 1993), characterized by a low energy index $\alpha$, a high energy index $\beta$, and a break energy. The EGRET instrument on the Compton Gamma Ray Observatory (CGRO) has provided several tantalizing observations at higher energies. GRB 940217 showed high energy emission more than an hour after the end of the lower energy emission, including one photon of 18 GeV (Hurley et al 1994). GRB941017 showed a high-energy component increasing as the lower energy emission faded (Gonzalez et al. 2003). One of the foremost scientific goals of *Fermi* is to fully explore the behavior of GRBs in the >1 MeV energy range.

### 1.2 *The Fermi Gamma-Ray Space Telescope*

The *Fermi* Gamma-Ray Space Telescope (Figure 1) was launched on June 11, 2008 into a 565 km orbit with an inclination of 25.6 degrees. The payload comprises two science instruments, the Large Area Telescope (LAT) and the Gamma-Ray Burst Monitor (GBM). The LAT observes gamma-rays above ~20 MeV from a wide variety of astronomical sources with unprecedented sensitivity. In particular, the LAT will make pioneering observations of gamma-ray bursts (GRB) at high energy (Atwood et al. 2009) The primary role of the GBM is to augment the science return from *Fermi* in the study of gamma-ray bursts by making observations at lower energies (~8 keV to ~40 MeV). The GBM-LAT combination thus provides burst spectra over seven decades in energy. Several GRBs have been detected by both LAT and GBM, including the particularly powerful burst GRB 080916C (Abdo et al. 2009).

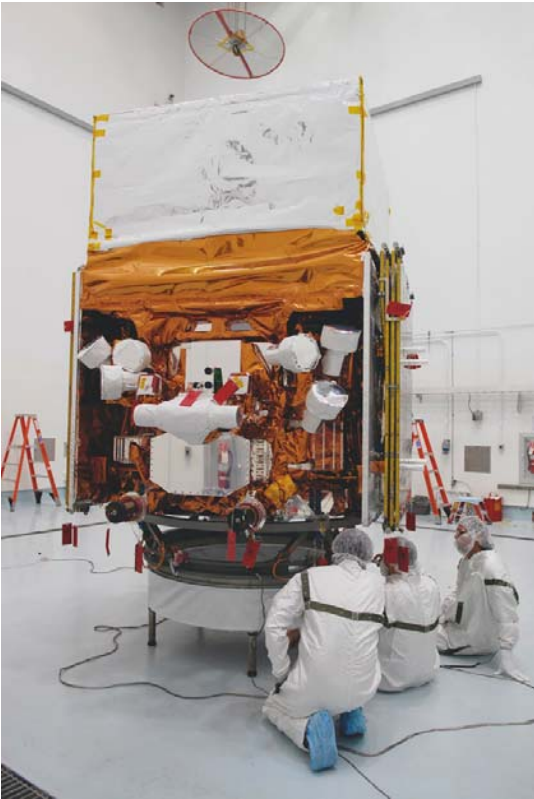

Figure 1: The *Fermi* Gamma-Ray Space Telescope being prepared for launch vehicle integration at Cape Canaveral. On top is the LAT with its reflective covering. Six of the GBM NaI detectors and one BGO detector can be seen on the side of the spacecraft. Photo credit: NASA/Kim Shiflett.

### 1.3 *GBM Science Overview*

The primary science goal of the GBM is the joint analysis of spectra and time histories of GRBs observed by both the GBM and the LAT. Secondary objectives are to provide near-real time burst locations on-board to permit repointing of the spacecraft to obtain LAT observations of delayed emission from bursts, and to disseminate burst locations rapidly to the community of ground-based observers. The GBM team will produce a catalog of GRBs containing spectral, temporal and location data.

Software on-board GBM detects and localizes bursts over the entire unocculted sky and transmits information to the LAT and to the ground in near real time. For particularly strong bursts, the spacecraft is re-oriented to allow bursts detected by GBM to be observed by the LAT for an extended time interval (nominally 5 hours). Since GBM will also trigger on solar flares, Terrestrial Gamma Flashes (TGFs) and Soft Gamma Repeaters (SGRs), a large amount of data are available for studies of these sources. When not processing a burst trigger, GBM transmits background data useful for a number of other studies, enabling a wide range of guest investigations. These data will be used to monitor variable X-ray sources using the earth occultation technique, as was done using the Burst and Transient Source Experiment (BATSE) on the Compton Gamma Ray Observatory (Harmon et al., 2002). Hard X-ray pulsars with periods greater than a few seconds will be monitored using Fourier transforms and epoch folding (Bildsten et al. 1997)

GBM is a collaboration involving scientists at the NASA Marshall Space Flight Center, the University of Alabama in Huntsville, the Max Planck Institute for Extraterrestrial Physics in Garching, Germany, and the Los Alamos National Laboratory. Overviews of GBM have been presented at several conferences (e.g., Lichti et al. 2007; Meegan et al. 2007; Meegan et al. 2008.)

1.4 *GBM Hardware Overview*

The GBM flight hardware comprises twelve thallium activated sodium iodide (NaI(Tl)) scintillation detectors, two bismuth germanate (BGO) scintillation detectors, a Data Processing Unit (DPU), and a Power Box. Figure 2 shows a functional block diagram of the GBM. The High Speed Science Data Bus (HSSDB) is the primary channel for sending GBM science data to the spacecraft for transmission to the ground. The Command and Telemetry Data Bus (CTDB) transmits commands from the spacecraft to GBM and housekeeping data from GBM to the spacecraft. The CTDB is also used to send immediate notifications of GRBs to the ground and for communications between the GBM and LAT, as described in Section 4. The Pulse per Second (PPS) signal provides a timing pulse to GBM every second. The Immediate trigger signal provides a prompt notification to the LAT that GBM has triggered.

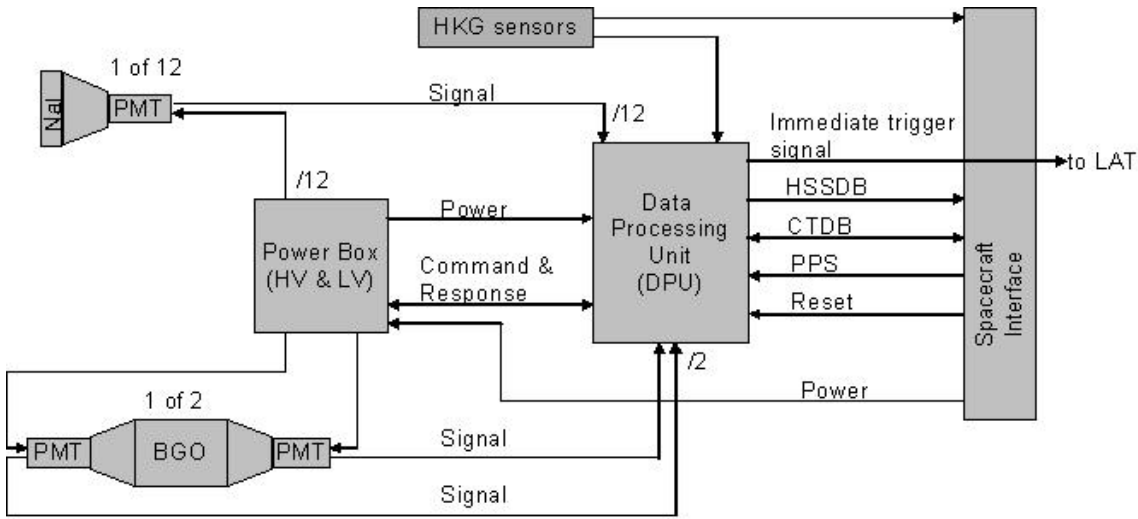

Figure 2. Functional Block diagram of GBM.

## 2. DETECTORS

### 2.1. *NaI(Tl) Detectors*

The NaI(Tl) detectors measure the low-energy spectrum (8 keV to 1 MeV) and are used to determine the directions to GRBs. The crystal disks have a diameter of 12.7 cm (5 in.) and a thickness of 1.27 cm (0.5 in.). They are packed in a hermetically sealed light-tight Aluminum housing (NaI is hygroscopic) with a 0.6 cm thick glass window (the glass is of the type NBK-7) with a diameter of 12.7 cm for the photomultiplier tube (PMT) attachment. The glass is glued to the Al housing with white Araldit. The entrance window is a 0.2 mm thick Be sheet for low-energy response. However, for mechanical reasons a 0.7 mm thick silicone layer had to be placed in front of the crystals as well which determines the low-energy threshold of 8 keV. In order to increase the light output the crystals are packed in a reflective white cover of Tetratec (on the front-window side) and Teflon (on the circumference). A photograph of a NaI detector flight unit is shown in Figure 3.

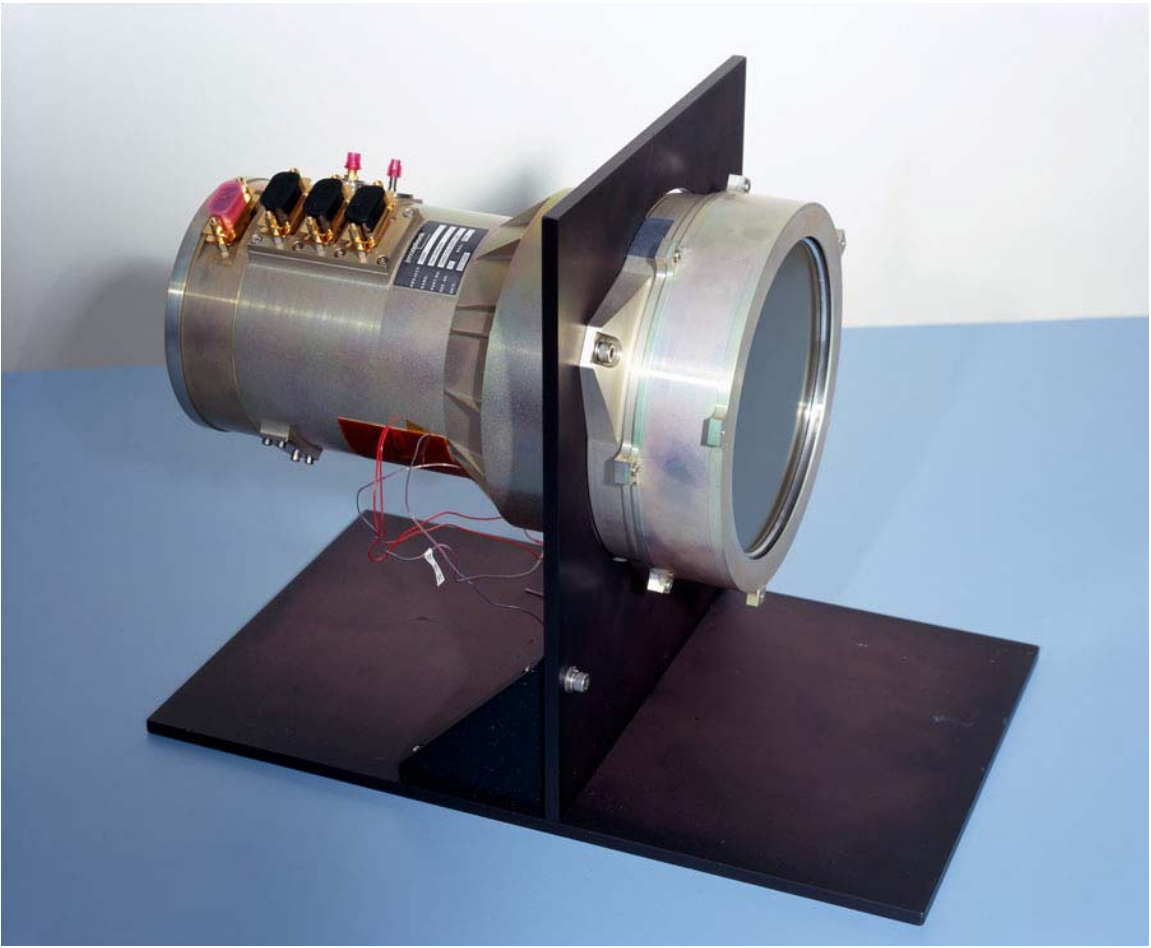

Figure 3. A NaI(Tl)-detector flight unit, consisting of a 5" by 0.5" NaI(Tl) crystal viewed by a PMT.

The axes of the NaI detectors are oriented such that the positions of GRBs can be derived from the measured relative counting rates, a technique previously employed by Konus and BATSE. A table of the direction angles of the NaI crystals in spacecraft coordinates are given in Table 1. The zenith angle is measured from the spacecraft +Z axis, (nominally aligned with the maximum effective area of the LAT), and the azimuth is measured clockwise from the +X (sun-facing) side of the spacecraft. The locations and orientations of the detectors are illustrated in Figure 4.

| Detector ID # | Azimuth (deg) | Zenith (deg.) |
|---|---|---|
| 0 | 45.9 | 20.6 |
| 1 | 45.1 | 45.3 |
| 2 | 58.4 | 90.2 |
| 3 | 314.9 | 45.2 |
| 4 | 303.2 | 90.3 |
| 5 | 3.4 | 89.8 |
| 6 | 224.9 | 20.4 |
| 7 | 224.6 | 46.2 |
| 8 | 236.6 | 90.0 |
| 9 | 135.2 | 45.6 |
| 10 | 123.7 | 90.4 |
| 11 | 183.7 | 90.3 |

Table 1. NaI detector measured orientations in spacecraft coordinates (see Fig. 4)

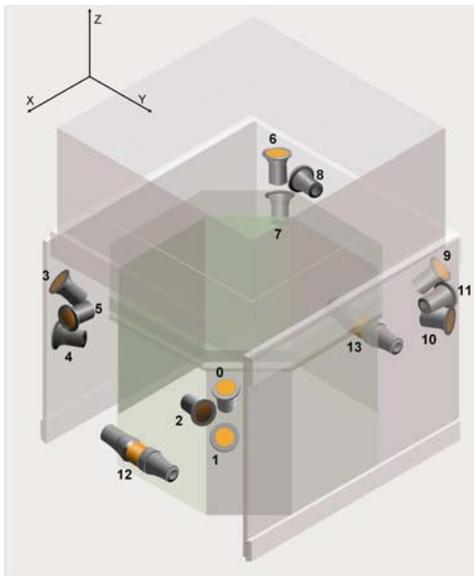

Figure 4. Locations and orientations of the GBM detectors.

## 2.2. BGO Detectors

The BGO detectors have an energy range of ~200 keV to ~40 MeV, overlapping at low energy with the NaI(Tl) detectors and at high energy with the LAT, thus providing for cross-calibration. They are positioned on opposite sides of the spacecraft so that any burst above the horizon will be visible to at least one of them.

The crystals have a diameter and a length of 12.7 cm (5 in.). The two circular glass side windows of the crystals are polished to mirror quality, while the cylindrical surface is roughened in order to guarantee a diffuse reflection of the generated photons. The crystals are packed in a carbon-fibre reinforced plastic (CFRP) housing that is held on both sides by Titanium rings. Titanium was to be chosen because its thermal expansion coefficient is close to that of BGO. The CFRP wrapping provides light tightness and guarantees the mechanical stability of the BGO units. The two rings serve also as holding structures for the two PMTs which are mounted on both sides to the crystals. The use of two PMTs results in better light collection and provides redundancy. The BGO detectors are mounted to a shell-type Titanium holding structure. A picture of a BGO detector is shown in Figure 5.

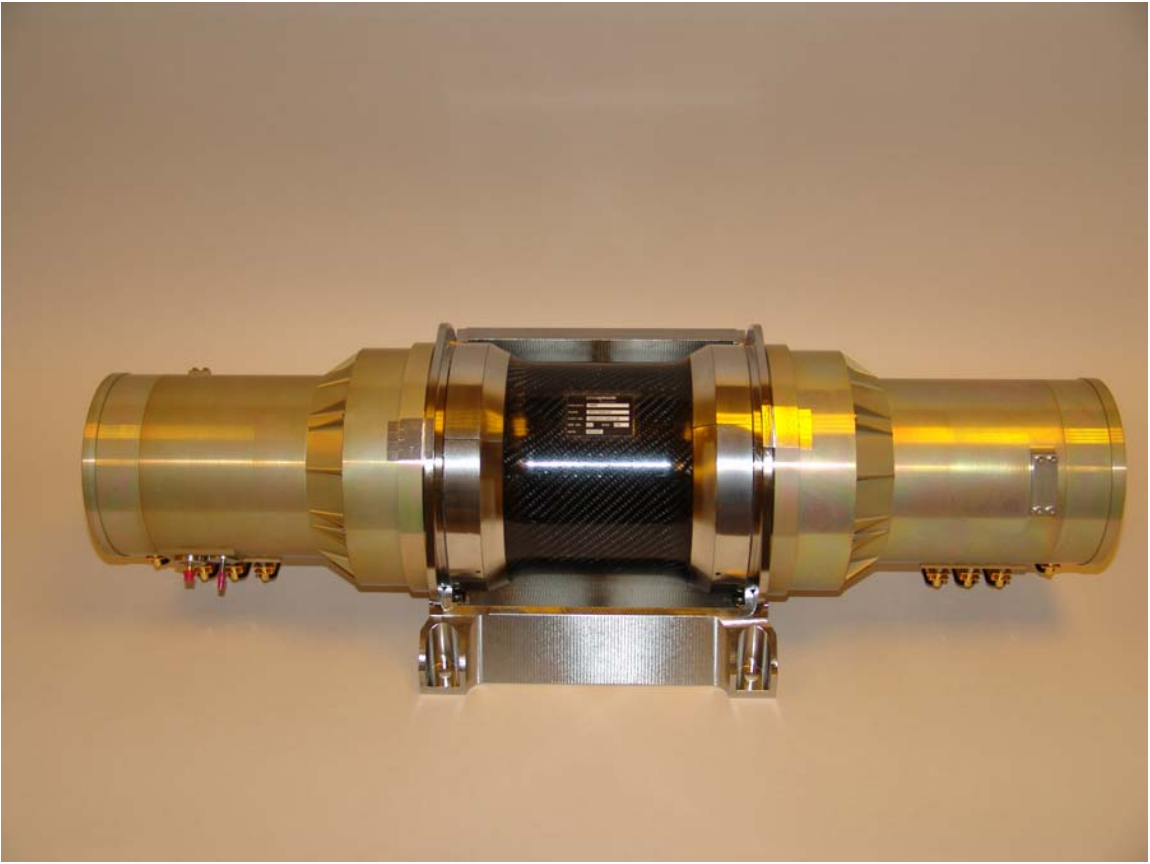

Figure 5. A BGO detector, consisting a 5" diameter by 5" thick bismuth germanate crystal viewed by two PMTs.

## 2.3. *Photomultiplier Tubes (PMTs) and Front-End Electronics (FEE)*

In order to transform the scintillation light into an electronic signal the crystals are viewed by PMTs. For both crystal types, PMTs of the type R877-MOD from Hamamatsu Corporation were used. These are 5 in. head-on 10-stage PMTs made from borosilicate glass with a bialkali (CsSb) photocathode and a box/grid dynode structure. The high voltage supplied to the PMTs is adjustable by command between 735 V and 1243 V in steps of 2 V.

The PMT housing includes a front-end electronics (FEE) board that shapes the PMT pulses (Figure 6). Signals from the PMT are fed to a charge sensitive amplifier (CSA) followed by a pulse shaper with base line restoration. This generates unipolar pulses that peak after ~0.37 μs. The voltage scale ranges from 0 V to 5 V, with 5 V corresponding to 1 MeV for the NaI detectors and to 40 MeV for the BGO detectors. Finally, a differential line driver (LD) produces bipolar outputs that are routed to the DPU through shielded twisted pair cables to eliminate common mode noise. The outputs for the two PMTs of each BGO detector and combined at the DPU.

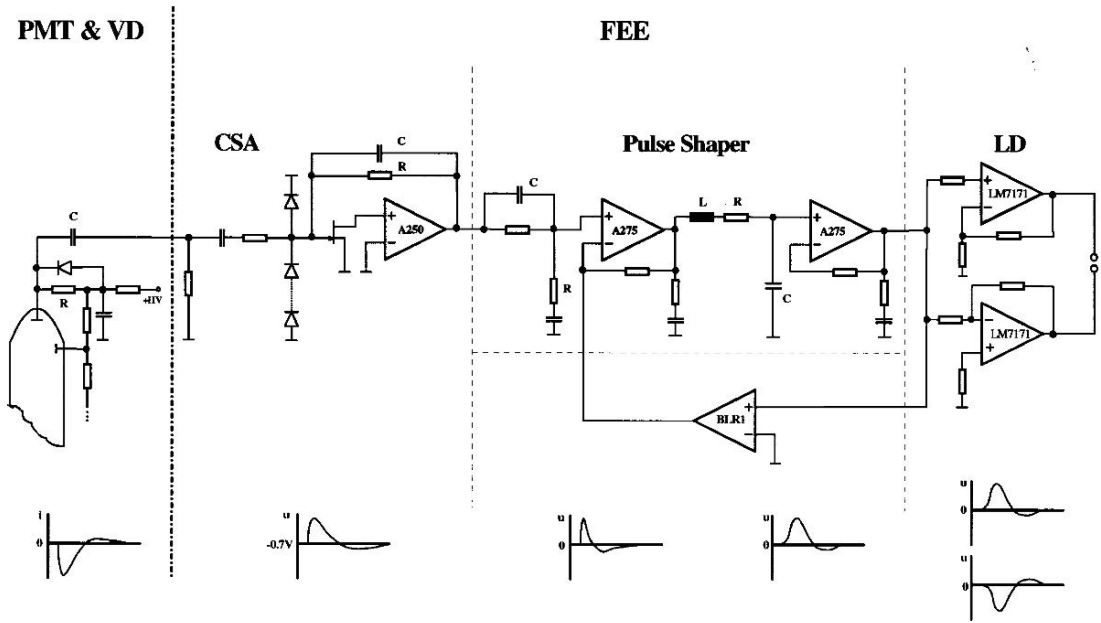

Figure 6: A functional block diagram of the front-end electronics (FEE) of each of the GBM photomultiplier tubes.

3. DATA SYSTEM

3.1. *Data Processing Unit Overview*

The Data Processing Unit (DPU) processes signals from the detectors, controls high and low voltage to the detectors, processes commands, and formats data for transmission to the spacecraft. Some functions, such as pulse height analysis and transmission of time-tagged event (TTE) data, are implemented in hardware, while others, such as GRB triggers, GRB localizations, and Automatic Gain Control, are implemented in software. All DPU functions are implemented on a single printed circuit board. For redundancy, there are two such boards, with fully cross-strapped inputs and outputs.

The DPU was designed, built, and tested by Southwest Research Institute of San Antonio, Texas, under a contract from Marshall Space Flight Center. The various DPU functions are described in more detail in the following sections.

3.2. *Pulse Height Analysis (PHA)*

The signals from each detector are sampled in the DPU by individual flash ADCs at 9.6 MHz and digitized into 4096 linear energy channels (Theis et al. 2006). If the level exceeds a programmable digital threshold, the peak of the pulse is detected by a firmware program running in a Field-Programmable Gate Array (FPGA). The peak is identified by requiring four successive lower samples (See Figure 7). An adjustable dead time after peak detection allows the bipolar signal to return to ground. The net dead time per event is nominally 2.6 μs. The peak heights are buffered and converted into 128-channel resolution (CSPEC) and 8-channel resolution (CTIME) using look-up tables that can be revised by command. The 8-channel edges are always chosen to coincide with 128-channel edges. Firmware in the DPU accumulates pulse height histograms for both 8-channel and 128-channel resolution. These are output over the spacecraft High Speed Science Data Bus (HSSDB) at regular intervals controlled by the GBM flight software as CTIME and CSPEC data, respectively. These data types are described further in section 3.4 below. A separate pair of histograms, referred to as ITIME and ISPEC, are read by the flight software every 64 ms and 61.44 s, respectively. ITIME is used to test for burst triggers. ISPEC is used to accumulate background spectra for Automatic Gain Control (AGC).

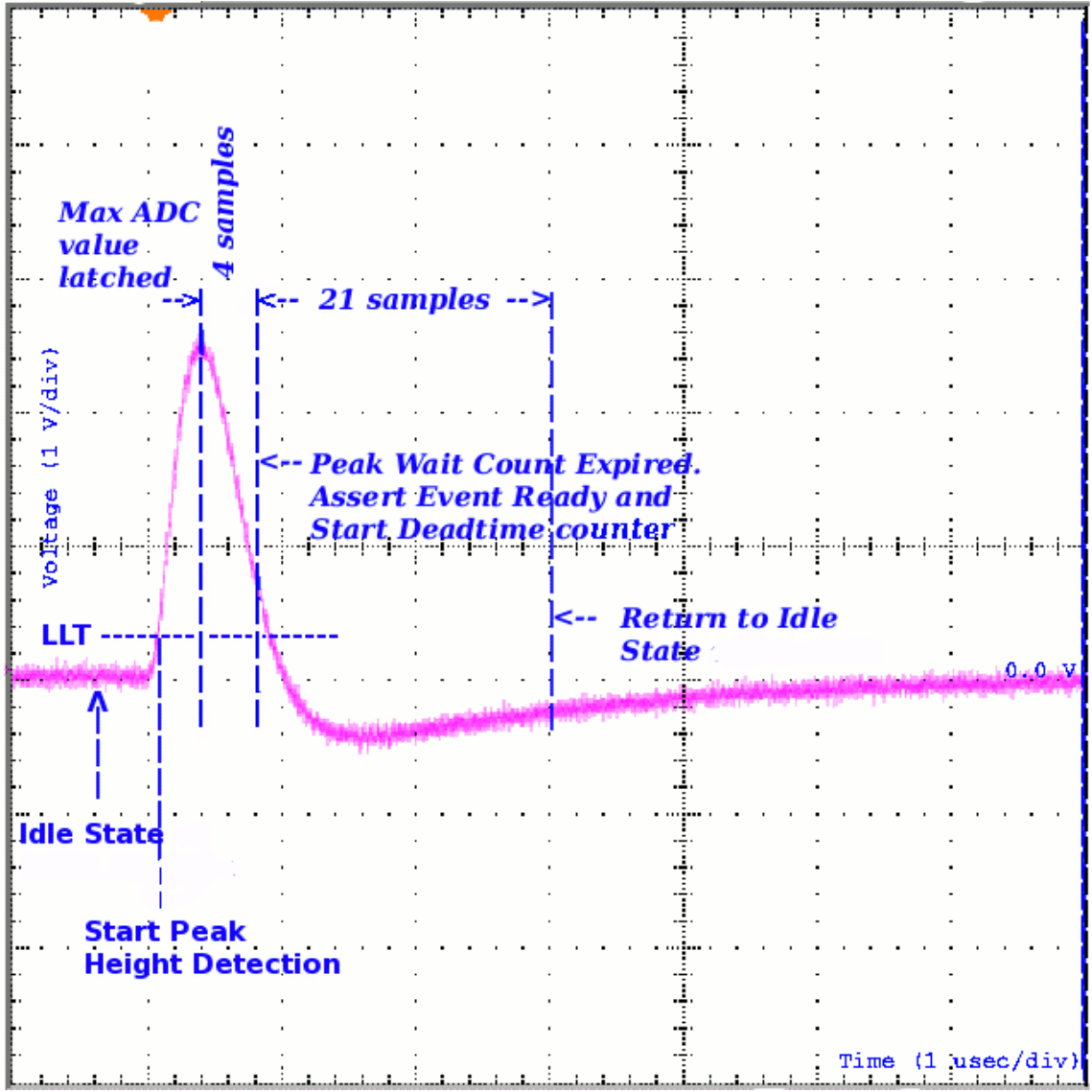

Figure 7: The shaped bipolar signal from a GBM detector PMT. The flash ADC in the DPU samples the signal (with a 12 bit resolution) at a constant rate of 9.6 MHz and a firmware program running in an FPGA identifies the peak sample at the 4$^{th}$ sample after the peak. The next 21 samples (this value can be changed by command) are ignored leading to a dead time of 2.6 μs.

### 3.3. *Flight Software*

The GBM flight software operates on an Atmel TSC695F/ERC32 microprocessor under the RTEMS real-time operating system. A boot version (version 1.2) of the flight software resides in fusible link programmable read-only memory (FPROM) with 128 kB capacity. This version provides a limited amount of science capability, but can accept commands and load flight software revisions. Version 1.2 itself can not be modified. After a fixed delay, this version initiates an automatic boot to a more advanced version (currently version 2.4) that resides in 512 kB of Electrically Erasable memory (EEPROM) in compressed form. This version can be modified in flight by command. The software uses 2 MB of RAM. All memory types include Error Detection and Correction (EDAC).

The flight software is interrupt driven. Interrupts are generated on the following events:
1. ISPEC data are available for reading (every 61.440 s).
2. ITIME data are available for reading (every 16 ms).
3. Attitude or time data are received from the spacecraft (6 times per s).
4. A ground command for GBM is received.
5. GBM is polled for housekeeping data.

The flight software performs the following tasks:
1. Configures the instrument for normal operation after boot-up.
2. Controls the high voltage to each of the PMTs, either in response to a command or for Automatic Gain Control (AGC).
3. Computes HV changes necessary for AGC using a line in the background spectrum.
4. Processes commands from the ground.
5. Triggers on detector rate increases that exceed threshold.
6. Controls production of Science Data, accelerating CSPEC and CTIME data during triggers, and producing TTE during triggers.
7. Calculates parameters for triggers, including flux, fluence, spectral hardness, and location.
8. Determines probability that a trigger is a GRB or other type of event.
9. Determines if a burst qualifies for a spacecraft repoint.
10. Sends burst data to the LAT over the Immediate Trigger Signal line and via spacecraft telecommand.
11. Maintains running average background rates and computes trigger thresholds.
12. Computes TRIGDAT data on triggers and transmits them to the spacecraft for downlink.
13. Turns off HV when the spacecraft is in the SAA.
14. Generates housekeeping telemetry.

### 3.4. *Data Types*

The DPU produces three types of science data packets: CTIME, CSPEC, and TTE.

The CTIME data consist of accumulated spectra from each detector with 8-channel pulse height resolution. The 8-channel pulse heights are derived from the internal 4096-channel (12-bit) pulse heights using look-up tables (LUT) controlled by the flight software. There are two CTIME look-up tables, one for the NaI detectors and one for the BGO detectors. The CTIME accumulation interval is controlled by the flight software. The range is from 64 to 1024 ms, in multiples of 64 ms, with a default value of 256 ms.

The CSPEC data consist of accumulated spectra from each detector with 128-channel pulse height resolution. The 128-channel pulse heights are derived from the internal 4096-channel (12-bit) pulse heights using look-up tables controlled by the flight software. There are two CSPEC look-up tables, one for the NaI detectors and one for the BGO detectors. The current LUTs are pseudo-logarithmic so that the spectral channel widths are commensurate with the detector resolution as a function of energy (Figures 8 and 9).

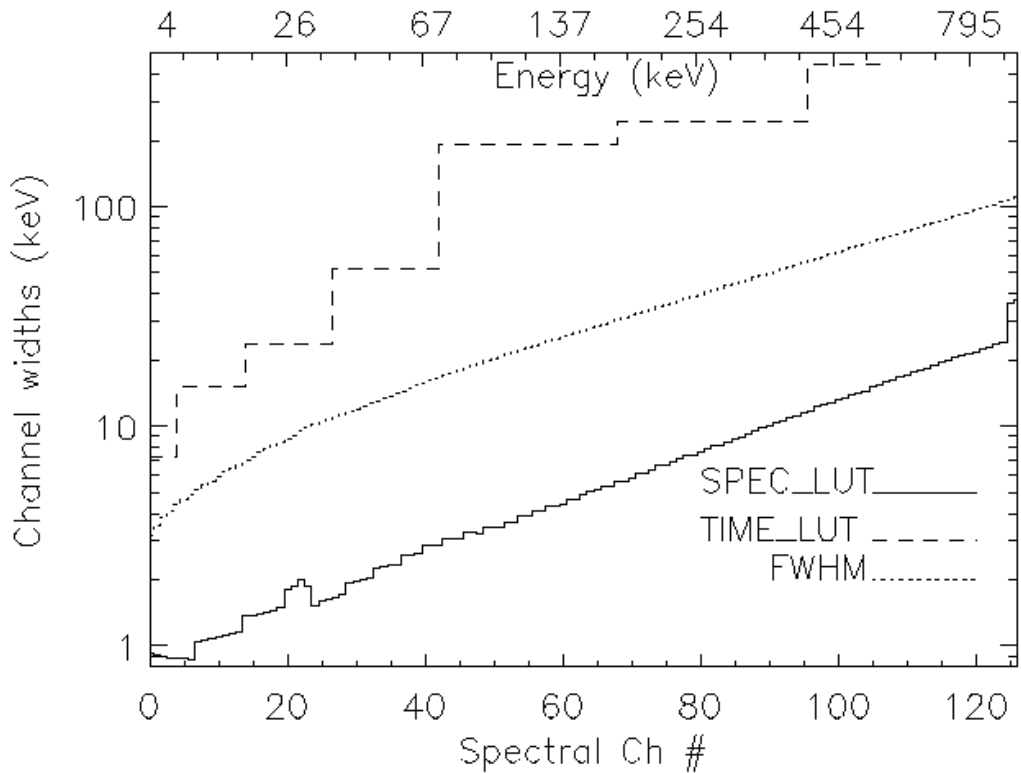

Figure 8: The CSPEC and CTIME Look-Up Tables for the NaI spectral channels in energy units plotted as function of spectral channel numbers. Also shown is the NaI energy resolution (FWHM) as a function of gamma-ray energy showing that a gamma-ray line is over sampled by a factor of ~4 in the CSPEC data.

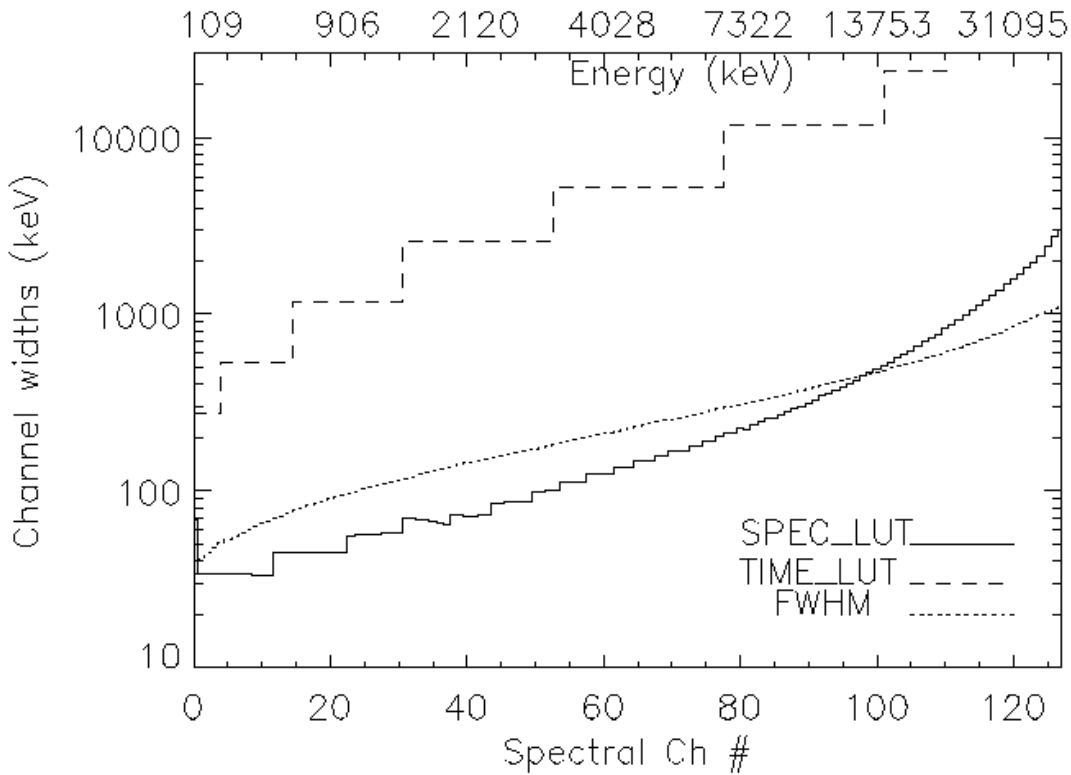

Figure 9: Same as figure 8 for BGO detectors. Here the energy resolution of the LUTs is sufficient to resolve spectral lines up to ~12 MeV.

The CSPEC accumulation interval is controlled by the flight software. The range is from 1.024 s to 32.768 s, in multiples of 1.024 s, with a default value of 4.096 s.

The TTE data consist of individually digitized pulse height events from the GBM detectors during bursts, and are encoded into 32-bit words. The pulse height is encoded into 7 bits using the same 128 channel boundaries as CSPEC. The detector identification is encoded into 4 bits. The time, modulo 0.1 s, is encoded into 16 bits with 2 µs resolution. A distinctive coarse time word is inserted into the data stream every 0.1 s to resolve ambiguities caused by rollover of the fine time data. TTE data are normally saved to a ring buffer with a capacity of 512k events. When a burst trigger occurs, the ring buffer is frozen and TTE data are routed directly to the spacecraft. The ring buffer is output later and provides ~30 seconds of pre-trigger data at typical background rates.

The absolute time is synchronized once per second using the spacecraft GPS time, which is accurate to ±1.5 µs. GBM sub-second time is derived from the 1.5 Mhz DPU clock, which drifts about 20 µs/s at typical on-orbit temperatures. For most applications, the ~20 µs maximum timing error is negligible. If greater accuracy is ever needed, the individual TTE times can be corrected for the clock drift. The timing agreement between LAT and GBM has been verified to be within 2 µs in ground testing using muons that traverse both the LAT detector and one of the GBM BGO detectors.

All the science data packets are output over the High Speed Science Data Bus (HSSDB), which is a parallel (8 bit) data bus with a maximum speed of 1.5 MB/s. Table 2 summarizes the GBM science data types.

| Name | Purpose | Temporal Resolution | Energy Resolution |
|---|---|---|---|
| CSPEC | Continuous high spectral resolution | Nominal: 4.096 seconds<br>During Bursts: 1.024 seconds<br>Adjustable Range: 1.024–32.768 s | 128 energy channels (adjustable channel boundaries) |
| CTIME | Continuous high time resolution | Nominal: 0.256 seconds<br>During Bursts: 0.064 seconds<br>Adjustable Range: 0.064–1.024 s | 8 energy channels (adjustable channel boundaries) |
| TTE | Time-tagged events during bursts | 2 microsecond time tags for 300 s after trigger; 500K events before trigger. Max. rate, all detectors: 375 kHz. | 128 energy channels (same as CSPEC) |

Table 2. Science Data Types

## 4. BURST OBSERVATIONS

### 4.1. *Triggers*

A burst trigger occurs when the flight software detects an increase in the count rates of two or more NaI detectors above an adjustable threshold specified in units of the standard deviation of the background rate. The background rate is an average rate accumulated over the previous $T$ seconds (nominally 17), excluding the most recent 4 seconds. Energy ranges are confined to combinations of the 8 channels of the CTIME data. Trigger timescales may be defined as any multiple of 16 ms up to 8.192 s. Except for the 16 ms timescale, all triggers include two phases offset by half of the accumulation time. A total of 120 different triggers can be specified, each with a distinct threshold.

The trigger algorithms currently implemented include four energy ranges: the BATSE standard 50 to 300 keV range, 25 to 50 keV to increase sensitivity for SGRs and GRBs with soft spectra, > 100 keV, and > 300 keV to increase sensitivity for hard GRBs and Terrestrial Gamma Flashes (TGFs). Ten timescales, from 16 ms to 8.192 s in steps of a factor 2, are implemented in the 50 to 300 keV range and the 25 to 50 keV range. The >100 keV trigger excludes the 8.192 s timescale, and the >300 keV trigger has only four timescales, from 16 ms to 128 ms.

When a burst trigger occurs, the flight software makes several changes to the data output. TTE data are re-routed from the pre-burst ring buffer to the spacecraft. Currently TTE data from all detectors are output, but the capability exists to reduce bandwidth by selecting only a subset of detectors. The CTIME and CSPEC integration times are decreased, nominally to 64 ms and 1.024 s, respectively. After a set time, nominally 300 s, the direct output of TTE data is terminated, and the pre-burst TTE buffer is dumped and restarted. Accelerated CTIME and CSPEC data rates continue for an additional time, nominally 600 s after the trigger.

At the nominal telemetry settings for CTIME and CSPEC data, GBM generates ~1.2 Gbits of data per day, plus a variable amount for each burst trigger. A weak burst generates ~0.3 Gbits of data, comprising mainly 300 s of background TTE data. A strong burst ($10^{-3}$ ergs cm$^{-2}$) generates ~0.5 Gbits of data.

### 4.2. *GRB Localization*

One of the goals of GBM is to provide information to allow reorienting the *Fermi* observatory to position strong bursts near the center of the LAT field of view (FoV) for extended observations. The GBM flight software contains algorithms to determine approximate locations of trigger events and to evaluate the probability that a trigger arises from a GRB.

When a burst trigger occurs, on-board software determines a direction to the source using the relative rates in the 12 NaI detectors. These rates are compared to a table of calculated relative rates for each of 1634 directions (~5 degree resolution) in spacecraft coordinates. The location with the best chi-squared fit is converted into right ascension and declination using spacecraft attitude information and transmitted to the ground as TRIGDAT data (described in section 4.5 below). The on-board table is calculated for rates integrated over 50 to 300 keV, for a specified assumed GRB spectrum. It includes nominal corrections for spacecraft scattering and atmospheric scattering, assuming the spacecraft +Z axis (the LAT axis) is zenith pointing. Normally the spacecraft +Z axis will be offset from the zenith by 35 degrees, which introduces a small error in the atmospheric scattering correction. Simulations indicate that this algorithm produces location errors of less than 15 degrees for strong bursts (fluence > 10 photons cm$^{-2}$). Location errors for weaker bursts are dominated by statistical fluctuations in the measured count rates.

Improved locations are automatically computed on the ground in near real-time by the Burst Alert Processor, described in section 6.4, and later interactively. The accuracy of GBM locations is addressed by Briggs et al. (2009).

### 4.3. *Spacecraft reorientations*

The *Fermi* Observatory incorporates the capability to autonomously alter the observing plan to slew to and maintain pointing at a GRB for a set period of time, nominally 5 hours, subject to earth limb constraints. This allows the LAT to observe delayed high energy emission, as has been previously observed by instruments on the Compton Gamma Ray Observatory (Hurley et al. 1994). Either the GBM or the LAT can generate an Autonomous Repoint Request (ARR) to point at a GRB. A request originating from GBM is transmitted to the LAT. The LAT either revises the recommendation, or forwards the request to the spacecraft. The LAT software may, for example, provide a better location to the spacecraft, or cancel the request due to operational constraints. The GBM flight software specifies different repoint criteria depending on whether or not the burst is already within the LAT FoV, defined as within 60° of the +Z axis. An ARR is generated by GBM if the trigger exceeds a specified threshold for peak flux or fluence. These thresholds are reduced if the burst spectrum exceeds a specified hardness ratio.

The GBM flight software includes an algorithm to classify triggers to avoid generating ARRs for non-GRB triggers. The probability that the trigger event is a GRB, as opposed to a solar flare, SGR, particle precipitation event, or known transient source, is calculated using a Bayesian approach that considers the event localization, spectral hardness, and the spacecraft geomagnetic latitude (McIlwain L coordinate).

The ARR criteria are adjusted to try to achieve a rate of about twice per month for repointing to a burst detected within the LAT FoV, and approximately twice per year for repointing to a burst not already in the LAT FoV. These criteria will be adjusted as the mission progresses so as to optimize burst observations by the LAT.

### 4.4. *Communication between LAT and GBM*

GBM transmits a variety of data on burst triggers to the LAT, which provides the capability to revise event filters to optimize GRB sensitivity, refine the location, or revise a GBM repoint recommendation. Every GBM trigger generates an Immediate Trigger Signal within 5 ms. A series of up to five calculated information packets are then sent, beginning 2 s after the trigger. These packets contain the trigger time, the event localization and categorization, and the timescale and energy band in which the trigger was generated. Finally, a single repoint request message is sent specifying whether or not the event meets the criteria for repointing the spacecraft.

The LAT also transmits data to the GBM whenever it produces a burst trigger. The GBM responds to such a signal by reducing the real-time telemetry rate to avoid conflicts with the LAT real-time data.

### 4.5. *TRIGDAT*

When a burst trigger occurs in either the LAT or the GBM, a real-time communication channel is opened which the two instruments share. Its function is to transmit information useful for rapid ground based observations. The information so transmitted by GBM is referred to as TRIGDAT data. The TRIGDAT data comprise information on background rates, the burst intensity, hardness ratio, and on-board localization and classification. Burst information is updated several times during the event. The bulk of the TRIGDAT data consists of a time history of the burst for all detectors with eight channel energy resolution and time resolution varying between 64 ms and 8 s.

## 5. PERFORMANCE

### 5.1. *Detectors*

The physical detector response of the GBM instrument (including the individual detectors and the influence of the *Fermi* spacecraft structure) is determined with the help of Monte Carlo simulations (see also Section 6.2), which are supported and verified by on-ground calibration measurements. The individual detectors were extensively calibrated and characterized at the Max Planck Institute for Extraterrestrial Physics (MPE) in 2005. All flight and spare detectors were irradiated with calibrated radioactive sources in the laboratory (from 14 keV to 4.43 MeV). The energy/channel-relations, the dependences of energy resolution and effective areas on the energy and the angular responses were measured (Bissaldi et al. 2009). Due to the low number of emission lines of radioactive sources below 100 keV, calibration measurements in the energy range from 10 keV to 60 keV were performed with the X-ray radiometry working group of the Physikalisch-Technische Bundesanstalt (PTB) at the BESSY synchrotron radiation facility, Berlin. These calibrations were useful for mapping the detector response in the region of the iodine k-edge at 33 keV. Calibrations with radioactive sources were also performed several times after the detectors were installed on the spacecraft.

The understanding of the instrumental response of the BGO detectors at energies above the energy range which is accessible by common laboratory radiation sources (< 4.43 MeV), is important, especially for the later cross-calibration with the LAT response in the overlap region between ~ 20 MeV to 40 MeV. In November 2006 the high-energy calibration of the GBM-BGO spare detector was performed using a small Van-de-Graaff accelerator at the Stanford Linear Accelerator Center (SLAC). High-energy gamma-rays from excited $^8$Be* (14.6 MeV and 17.5 MeV) and $^{16}$O* (6.1 MeV) were generated through (p, γ)-reactions by irradiating a LiF-target with ~400 keV protons.

Light emitted by an inorganic crystal is generally proportional to the energy deposited by the electron for energy greater than ~400 keV, but shows significant non-linearity at lower energy (Leo, W.R. 1992). A prominent feature at 33 keV arises from the K-edge in iodine. A similar behavior was observed with the BATSE Spectroscopy detectors (SD; Band et al., 1992).

The GBM detector calibrations are described in detail by Bissaldi et al. (2009). Here we summarize some of the major results. The energy resolution versus photon energy is shown in Fig. 10 for both an NaI and a BGO detector. The displayed data for the NaI detector excludes the BESSY results because the BESSY X-ray beam irradiated only a ~1 cm$^2$ area of the entrance window, which is not representative of the response for full-disk irradiation. The dependence of the BGO detector energy resolution includes the values of lines recorded with radioactive sources and the width of the 17.5 MeV and 6.1 MeV line. The resolution was determined by accounting for the 511 keV single escape peaks for lines above 4 MeV. The 14.6 MeV line is not included because it is intrinsically broadened. The curves are fits using the formula $FWHM^2 = bE + cE^2$, where the average values of *b* and *c* are 0.9 and 0.1 for NaI detectors, and 3.0 and 0.036 respectively for BGO detectors.

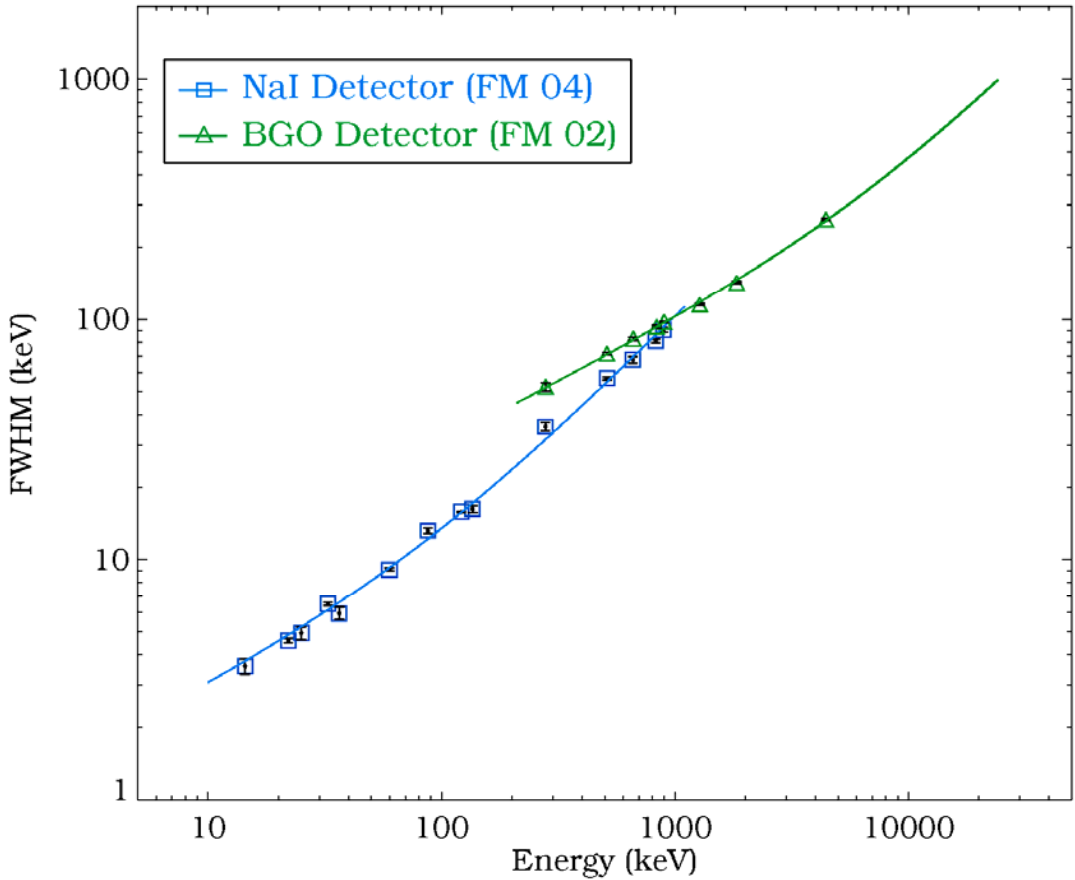

Figure 10. Dependence of the detector energy resolution for a NaI detector (squares) and a BGO detector (triangles).

The 14.4 keV gamma-ray line from a calibrated (3%) $^{57}$Co-source and the $K_\alpha/K_\beta$ X-ray lines of $^{109}$Cd at 22.1 keV and 25 keV and $^{137}$Cs at 32.06 and 36.06 keV were used to determine the transmissivity and effective area at low energies. The dependence of the effective area on energy, determined at normal incidence, is shown in Fig. 11 for both detector types in comparison with the simulated response. Agreement between laboratory measurements and simulations (upper panel) is better than ±5%, averaged over all detectors (Hoover et al. 2009). This is the estimated systematic uncertainty in the simulated response incorporated into the shaded curves of the on-orbit simulations of the lower panel. The on-orbit simulations include the effects of spacecraft blockage and scattering. In some cases the total effective area above a few hundred keV is significantly increased due to photons that scatter from the spacecraft into the detector.

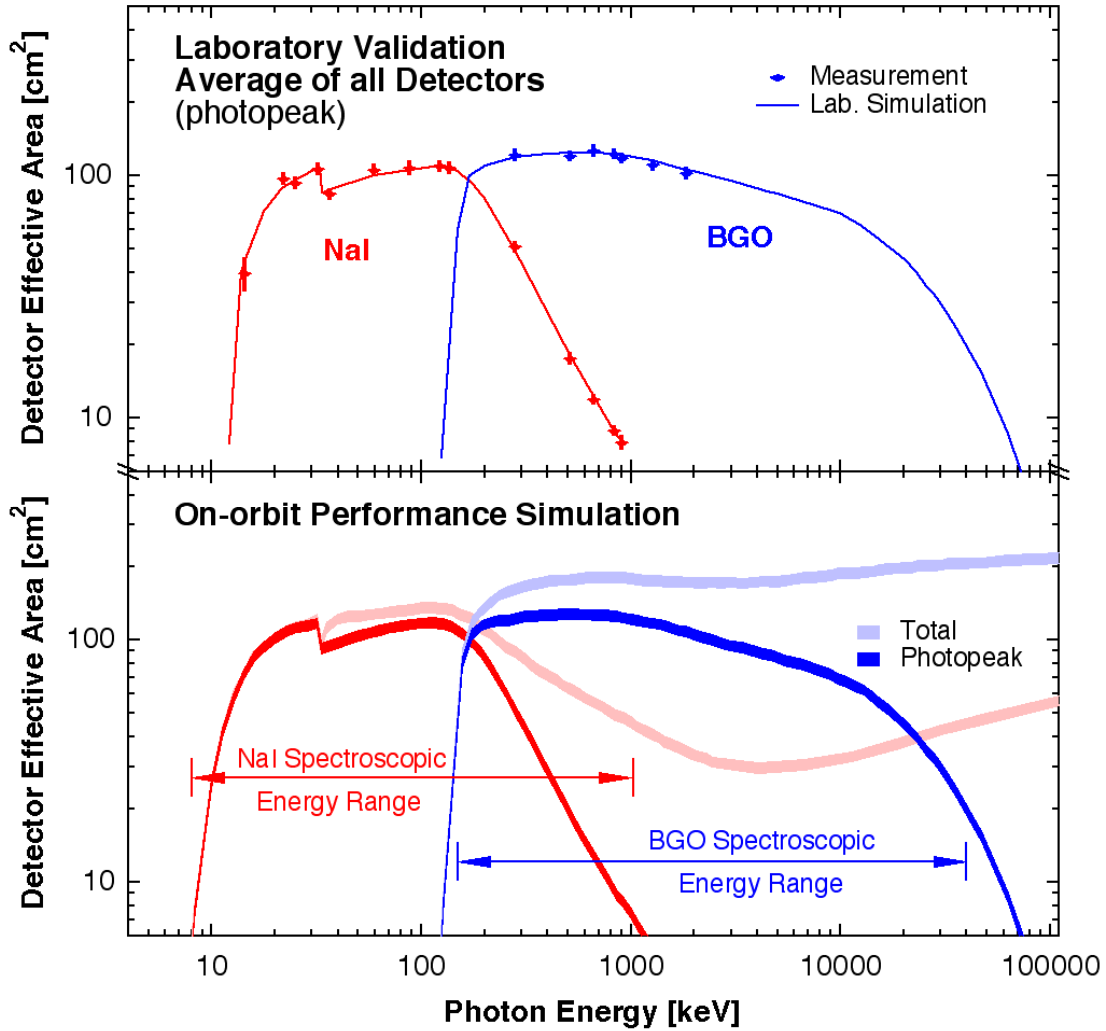

Figure 11. Energy dependence of the effective area at normal incidence, for both detector types. The lower panel includes the simulated effects of the spacecraft for a representative detector.

The angular dependence of the NaI detector response (Figure 12) and BGO-detector response (Figure 13) show shadowing effects of absorbing materials (e.g., detector housings, PMTs and FEE). In the case of the NaI detector the angular response for the flat crystal is approximately cosine, but differs for low energies (32.89 keV) and high energies (661.66 keV), since at these energies the dependence of the transmissivity on the angle of incidence is more important than at lower energies.

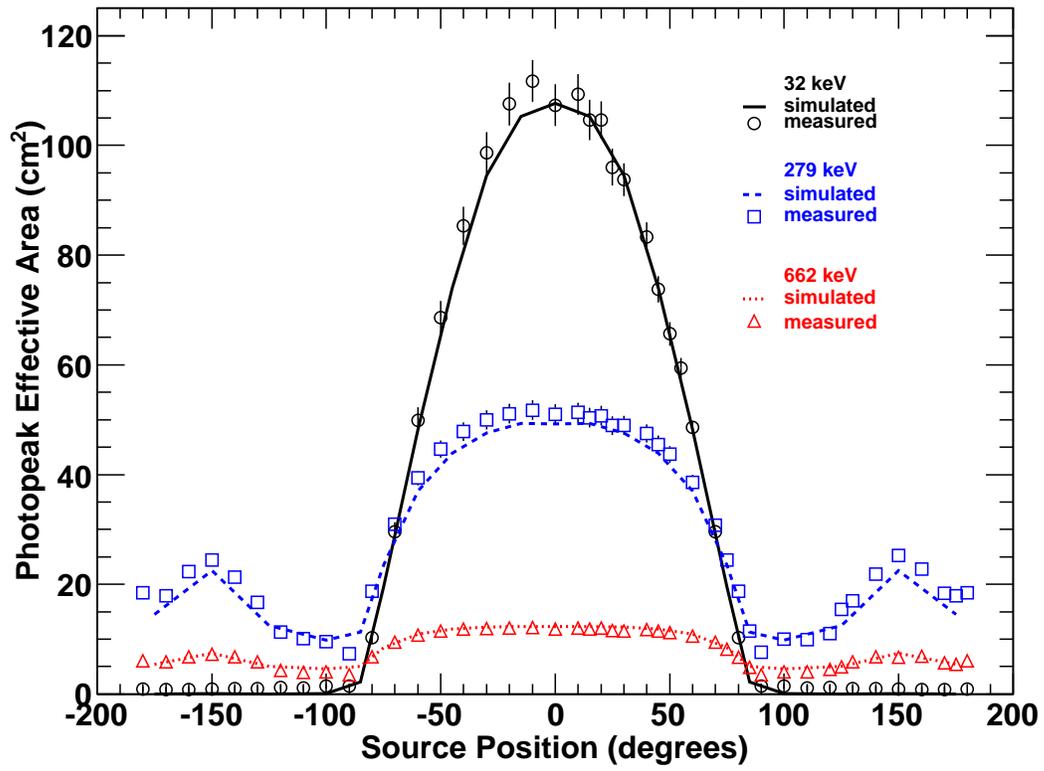

Figure 12. Angular dependence of the NaI detector effective area.

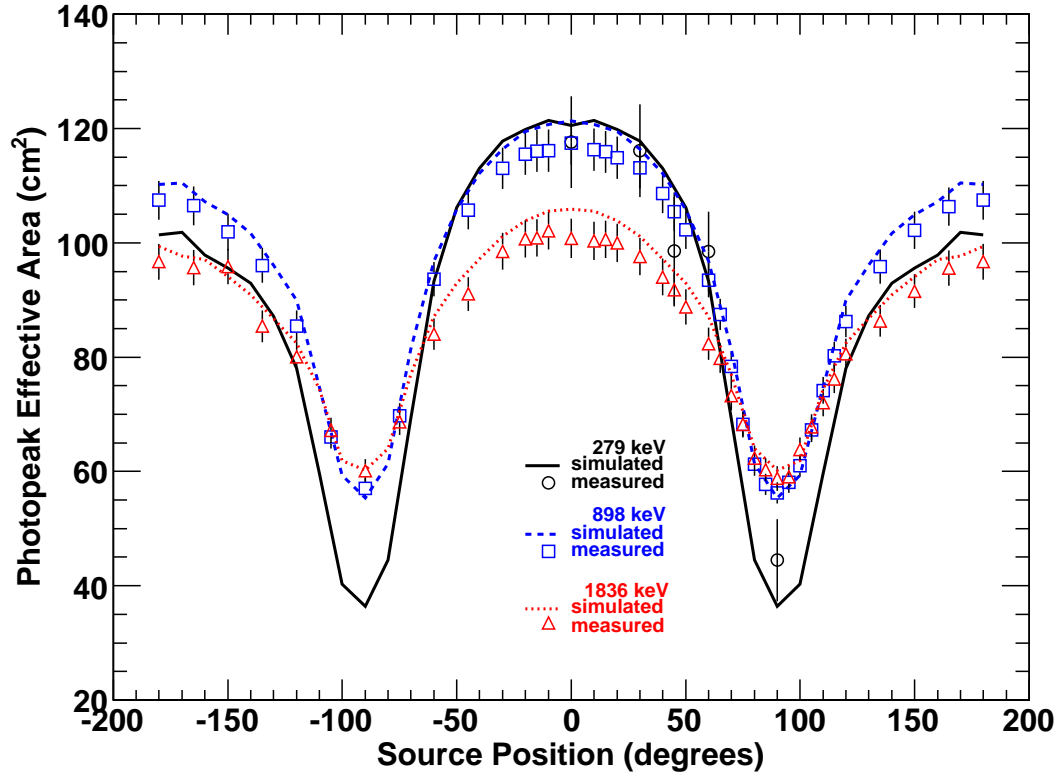

Figure 13. Angular dependence of the BGO detector effective area.

### 5.2. *High Rate Performance*

Two effects typically impair performance of scintillation detectors at high photon rates: dead time and pulse pile-up. Dead time limits the maximum rate of digitized pulses. The nominal dead time setting for GBM results in a fixed dead time of 2.6 μs per event, independent of energy, except that events that fall into the last (overflow) channel (4095) are always assigned 10 μs dead time. The effective dead time is a mean of these two values weighted by the total number of events in spectral channels other than the overflow channel and those in the overflow channel respectively. This adds a weak spectral dependence to the effective dead time of GBM.

Pulse pile-up occurs when the count rate is so high that the pulses from successive events overlap in the detector front-end electronics. This causes distortions in the measured spectrum that are difficult to characterize. Studies were performed to evaluate the effects of pulse pile-up in GBM. The effects were quantified in terms of systematic errors in the determination of the power law slope and peak energy of a burst spectrum characterized by a Band function (Band et al. 1993) with peak energy of 200 keV and a high-energy index of -2.15 (figure 14). It was found that up to rates of 50,000 cps, the peak energy has an error less than 1.5%, and the power law index has an error less than 0.6%.

Gamma-ray bursts with peak fluxes of this magnitude in individual detectors are expected to occur less than once per year.

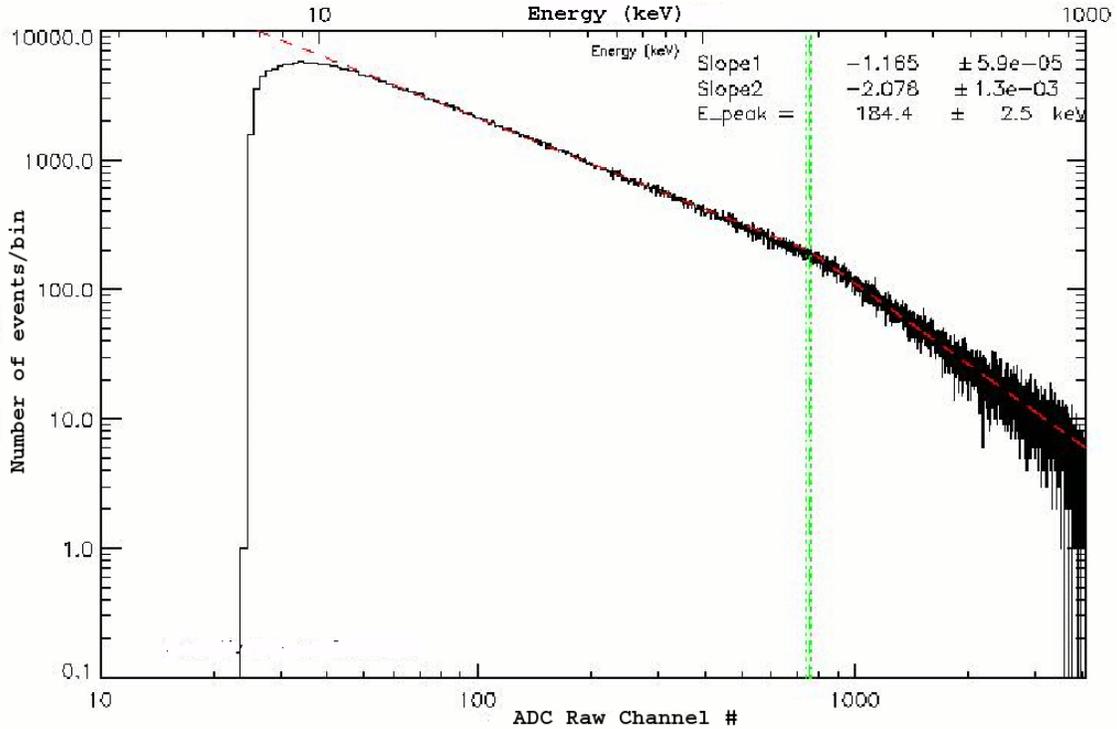

Figure 14: Pulse pile-up effect in the GBM DPU at a detector count rate of 100 kHz. The simulated band spectrum has input parameters of $\alpha = 1.24$, $\beta = 2.15$ and $E_{peak} = 196.7$ keV. The output spectrum when fitted has $\alpha = 1.17$, $\beta = 2.08$ and $E_{peak} = 184.4$ keV.

In GBM, there is an additional problem at high rates due to the limiting data rate of the High Speed Science Data Bus. During an intense trigger the total TTE count rate from all detectors could exceed the HSSDB limit of 1.5 MB/s (= 375k events/second). In such a case, TTE events are clipped at this maximum rate, resulting in an irrecoverable loss of data. We have seen such effects during intense SGR triggers, but not in any GRBs so far. Very short events such as Terrestrial Gamma-ray Flashes (TGFs) are unaffected, since they are completely contained within the pre-burst TTE ring buffer.

### 5.3. On-orbit Background

Spectral and temporal properties of the background in the NaI and BGO detectors are shown in Figures 15 through 18. Figures 15 and 16 show 2.4 hr. accumulations of background spectra from a NaI and a BGO detector. Figures 17 and 18 show background rates (1.024 s accumulations) over a one day time interval for a NaI and a BGO detector.

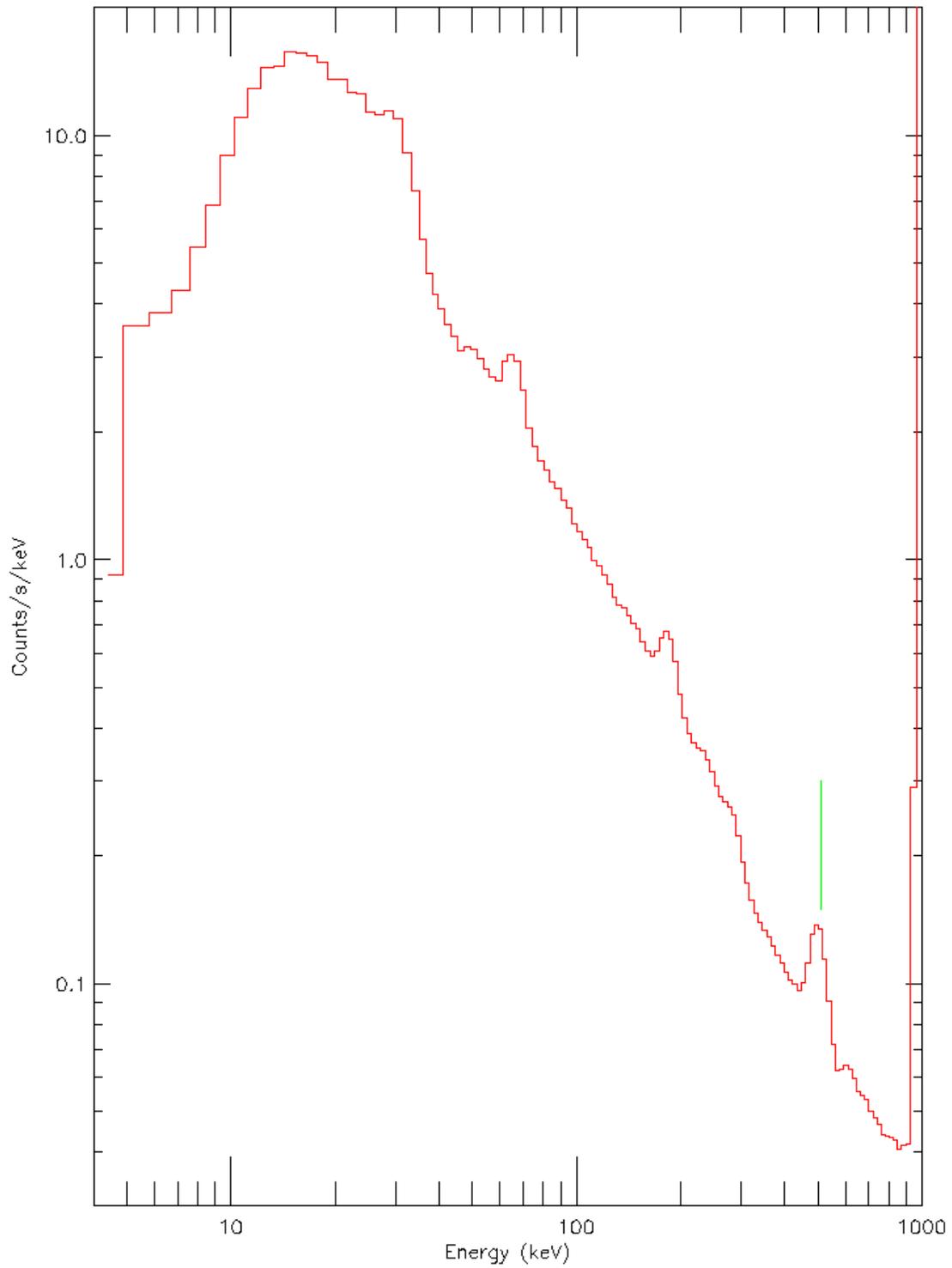

Figure 15: Background spectrum for a NaI detector (2.4 hour accumulation). The 511 keV line used for AGC is marked with a vertical line. Other features are described in the text.

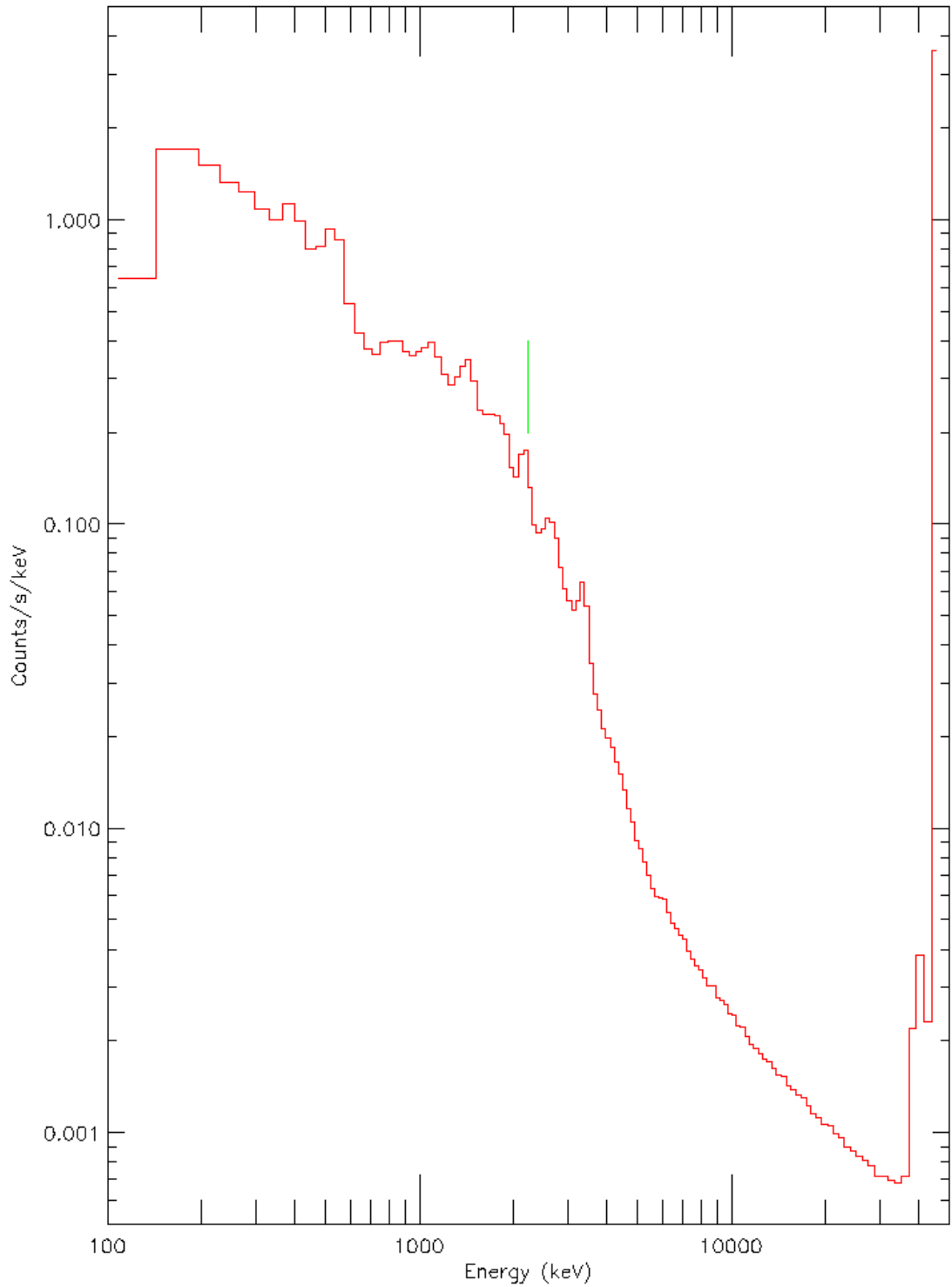

Fig 16: Background spectrum for a BGO detector (2.4 hour accumulation). The 2.2 MeV line used for AGC is marked with a vertical line. Other features are described in the text.

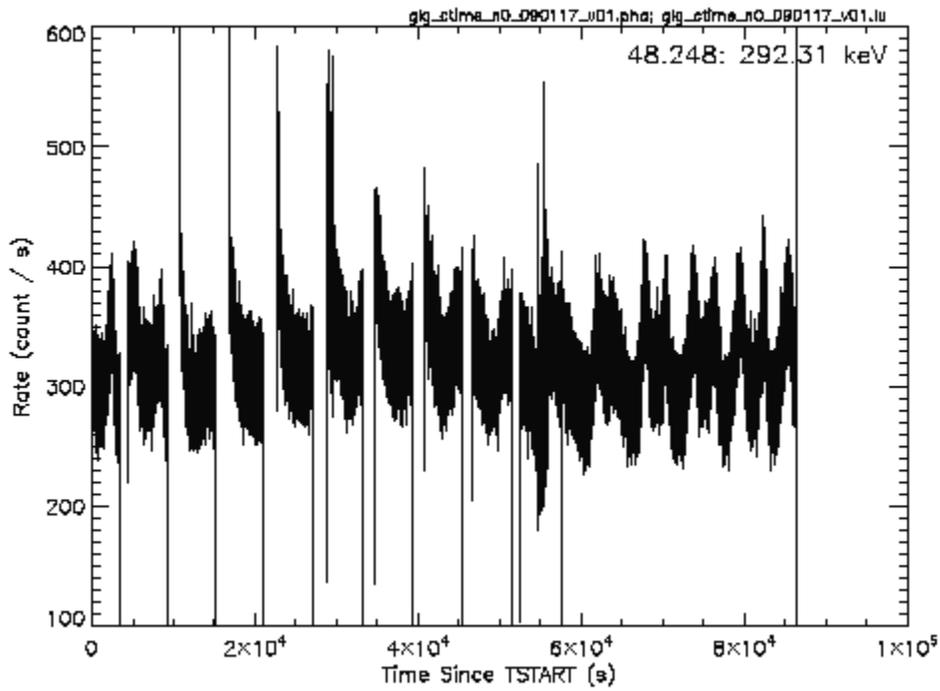

Figure 17. NaI detector background rates in the 50-300 keV energy range.

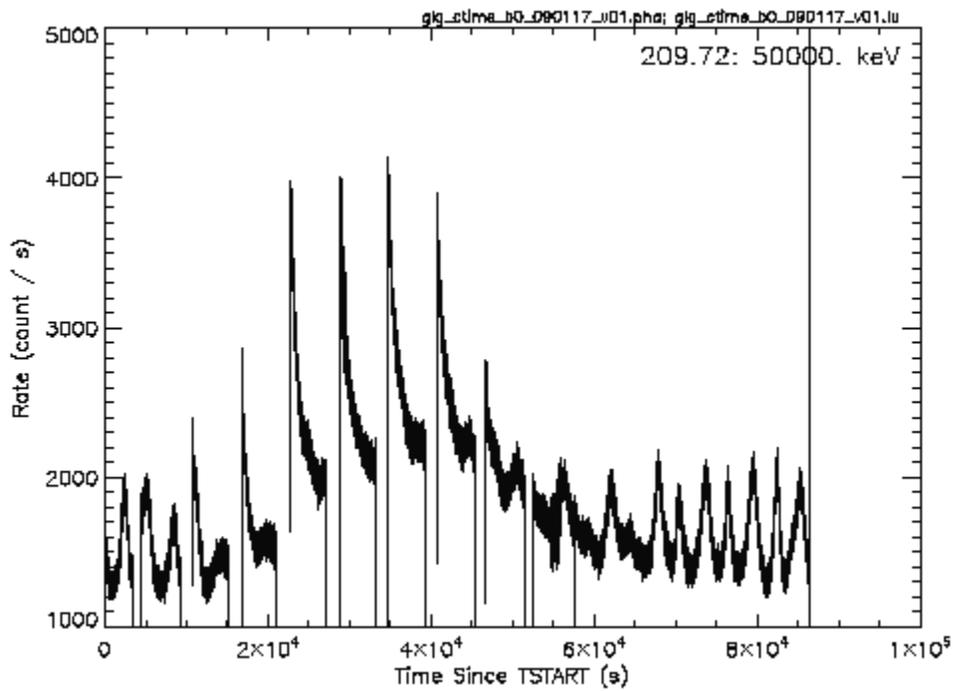

Figure 18. BGO detector background rates over the full BGO energy range.

The underlying background continua above ~150 keV in both types of detectors are dominated by secondary cosmic-ray-produced photons. This background source has as a major component the Earth gamma-ray albedo, and to a lesser extent, secondary gamma rays generated in local materials of the detectors and the spacecraft. This component of the background is modulated by the geomagnetic latitude, as the spacecraft traverses its orbit (see Figures 17 and 18). Below ~150 keV, the uncollimated GBM detectors have a significant counting rate from the diffuse x-ray background. This contribution is, of course, dependent upon the sky viewing fraction of the detectors that is not blocked by the Earth or spacecraft. Extensive preflight calculations were made to estimate the background contributions from all of the above sources.

In the NaI detectors, passive materials in front of the NaI detectors begin to limit the response of the detectors significantly, in the energy region ~8 keV to ~20keV, resulting in the low-energy drop seen in Figure 15.

The highest few channels of the spectra from both types of detectors are the "overflow channels". These channels contain counts due to energy deposits in the scintillators that are above the maximum set for the detector, ~1 MeV for the NaI detectors and ~45 MeV for the BGO detectors. The gradual upturn of the background in the BGO detectors above ~10 MeV is due to the logarithmically increasing width of the energy channels, increasing faster than the spectral decrease of the gamma-ray background continuum.

In Figure 15, the 511 keV line from positron annihilation in the atmosphere and nearby materials is clearly seen. This line is used for the automatic gain control (AGC) system, described in section 5.4. Two other prominent lines are observed from two excited energy levels of $^{127}$I, which are at 57.6 keV and 202.9 keV. Another blended line feature between 600 and 700 keV also likely arises from a grouping at excited states of $^{127}$I in this energy range. All of these features are believed to be due to fast neutron excitation, followed by nuclear de-excitation of this isotope.

The background spectrum of a BGO detector, shown in Figure 16, also shows line features. The strong line at 2.2 MeV is due to neutron capture in the large amount of hydrogen contained in the hydrazine tanks of the spacecraft. This line is used for the AGC system of the BGO detectors. A line at 1.46 MeV is due to $^{40}$K, primarily from the potassium contained in the glass in the PMTs of the BGO detectors. Nitrogen and oxygen nuclear excitation lines also appear but are not well-separated from the continuum. Other line features are mainly due to unresolved and/or unidentified activation and excitation lines in the BGO crystal by protons and neutrons in the ambient environment.

In the temporal plots (Figures 17 and 18), times of zero rate are due to turning off the PMTs during South Atlantic Anomaly (SAA) passes. These plots show the effect of activation by the SAA, particularly in the BGO detectors, as well as high rates near the SAA boundaries. The NaI rates are shown for the primary trigger energy range of 50 to 300 keV, and average ~320 cps, with very little variation among the 12 detectors.

### 5.4. *AGC & Gain Stability*

The following effects can produce changes in the detector gains: temperature changes of the detectors and the HV power supply, variations in the magnetic field at the PMT, and PMT aging. The GBM flight software compensates for long timescale gain changes by adjusting the PMT high voltage to keep a background line at a specified energy channel. The 511 keV annihilation line is used for the NaI detectors and the 2.2 MeV line from neutron capture on hydrogen is used for the BGO detectors. Spectral accumulations of 90 minutes are used to attain the required statistical accuracy. This Automatic Gain Control (AGC) technique is very similar to that successfully employed with BATSE. The AGC algorithm allows small drifts in the gains because of the discrete voltage steps and to allow for statistical fluctuations in the measurements. The high voltage is adjusted when the fitted spectral line has moved from the desired position by more than 1.4% in the NaIs and by more than 1.0% in the NaIs. The gains of the NaI detectors are typically adjusted two or more times per day, showing a daily period. The gains of the BGO detectors, which have much higher thermal inertia, are adjusted much less frequently.

Temperature changes and magnetic field changes can produce small gain variations on less than orbital time scales, too rapid to be removed by the AGC. The magnetic field variation is due not only to the changing intensity and direction of the Earth's field, but also to the operation of the spacecraft's magnetic torque bars. An analysis of these effects was performed assuming worst case generation of torque bar fields and worst case thermal model. The predicted RMS gain variations over one orbit are <2% for any detector.

### 5.4. *On-Orbit Trigger Performance*

The GRB trigger threshold is defined as the weakest burst with a 50% probability over the unocculted sky for producing an on-board trigger. The intensity is specified in terms of peak flux over 1 second in the 50-300 keV energy range. For the observed background rate of 315 cps (see Figure 17) and a significance threshold of 4.5$\sigma$, the trigger threshold is 0.74 photons-cm$^{-2}$-s$^{-1}$.

Burst triggering was enabled on July 11, 2008. There have been 404 triggers between then and March 31, 2009. Table 3 lists the sources of these triggers as determined by ground analysis (not necessarily the classification determined by the flight software). The class "other" includes particle precipitation events, accidentals caused by statistical fluctuations in the background, Cygnus X-1 fluctuations, and events with uncertain classifications. The requirement that at least two detectors exceed threshold effectively eliminates triggers from phosphorescence spikes caused by high-Z particles (Fishman & Austin 1977, Kouveliotou et al. 1992). Nearly 90% of the triggers are scientifically interesting. The GRB trigger rate is ~260 bursts/year. The average on-board location error for GRBs with precisely known locations is 9 degrees, consistent with pre-launch predictions.

| Trigger Classification | Number of Triggers |
|---|---|
| Gamma-ray burst | 183 |
| SGR 1547-5488 | 124 |
| SGR 1501+4516 | 27 |
| SGR 1806-20 | 2 |
| AXP 1E1547.0-5408 | 14 |
| Solar Flare | 1 |
| Terrestrial Gamma Flash | 8 |
| Other | 45 |
| TOTAL | 404 |

Table 3. Burst trigger statistics for the period July 11, 2008 to March 31, 2008.

## 6. OPERATIONS AND DATA ANALYSIS

### 6.1. *GBM Instrument Operations Center*

The primary GBM Instrument Operations Center (GIOC) is implemented at the National Space Science and Technology Center (NSSTC) in Huntsville, Alabama. The GIOC autonomously receives GBM data from the *Fermi* Mission Operations Center (MOC) and prepares and transmits higher level data to the *Fermi* Science Support Center (FSSC). The GIOC also maintains operations planning tools and generates command loads for GBM. An operations center at the Max Planck Institute for Extraterrestrial Physics (MGIOC) is functionally equivalent to the GIOC except for the capability to generate command loads. A backup GIOC is implemented at the FSSC in case the GIOC becomes inoperable for more than 24 hours.

Daily science operations consist mainly of processing burst triggers. This includes refined localizations and classifications, computation of GRB duration, peak flux, fluence, and spectral parameters, and submission of GCN circulars. These duties are performed by an assigned Burst Advocate (BA), alternating between GIOC and MGIOC personnel on 12 hour shifts.

### 6.2. *Detector Response Matrices (DRMs)*

Analysis of GBM data products is fundamentally a process of hypothesis testing wherein trial source spectra and locations are converted to predicted detector count histograms, and these are statistically compared to observed data. The key element in the conversion process is detailed and accurate representation of the composite GBM instrument response function. This is captured in the form of Detector Response Matrices (DRMs) for all individual GBM detectors. The DRMs, which contain the multivariate effective detection area, include the effects of angular dependence of the detector efficiency, partial energy deposition in the detector, energy dispersion and nonlinearity of the detector, and atmospheric and spacecraft scattering (and shadowing) of photons into the detector. They are therefore functions of photon energy, measured (deposited) energy, the direction to the source with respect to the spacecraft, and the orientation of the spacecraft with respect to the Earth.

The DRMs are generated using the General Response Simulation System (GRESS), a simulation and modeling code based on the GEANT4 Monte Carlo radiation transport simulation toolkit (Agostinelli et al. 2003). The GRESS code and models were extensively validated against data from radioactive source calibration of individual GBM detectors as well as data from a radioactive source survey of the integrated *Fermi* spacecraft (see Hoover et al. 2009). After iterative validation analysis, data from the detector-level calibrations (as a function of angle and energy) performed at MPE agree with GRESS results to within better than ±5% for the BGO detectors, and ±10% for the NaI detectors. Most of the remaining discrepancy is a by-product of experimental uncertainty and uncalibrated detector-to-detector variations.

The *Fermi* source survey provided the only means to validate the GRESS spacecraft model used to compute important scattering and absorption effects between the spacecraft and the GBM detectors. Two collimated radioactive sources ($^{137}$Cs and $^{60}$Co) were each used to illuminate the spacecraft from 12 different directions. The full configuration of the survey was modeled with GRESS, and comparisons between simulations and measured data were used to iteratively improve the spacecraft model. With the resulting model, photopeak efficiencies for detectors with angles <90 deg from the source agree to better than 5% RMS. At larger angles, where major parts of the spacecraft shadow the detectors, the RMS error is significantly larger (~25%). This is the unavoidable tradeoff between simulation speed and the ability to accurately model small details of the spacecraft. Fortunately, the most valuable GBM observational data always come from detectors with source viewing angles < 90 deg.

In practice, the multivariate GBM DRMs are separated into two components for GRESS computation efficiency. The first component includes the energy and angular dependent response of detectors with the *Fermi* spacecraft. It is stored in a large data set called the direct response database. The second component includes the effects of photons scattering in Earth's atmosphere as a function of energy and source-Earth-spacecraft geometry. It is stored in a large data set called the atmospheric response database. In the data analysis process, these two components are combined together for a specific set of observing conditions to form the composite set of DRMs. This process is similar to that used for the analysis of BATSE data, with the exception that for BATSE the DRMs were further separated into detector effects and spacecraft effects (Pendleton, G.N. et al., 1995). Advances in computing allow us to combine these important effects for GBM, and thereby preserve more detector response information. A set of composite DRMs is provided as a standard data product for each triggered GRB as described below.

### 6.3. *Data Products*

Raw data are provided by the spacecraft telemetry to the ground and are processed by the MOC. Level 0 data undergo minimal processing: no information is lost, but duplicate data packets are removed, quality checks are made, and the data packets are time-ordered. Performed at the MOC, Level 0 processing converts the raw telemetry into the Level 0 data.

Level 1 data are generated by automated pipeline processing of Level 0 data, and serve as the starting point for scientific analyses by the user community and the GBM instrument team. Level 1 processing of GBM data is performed at the GIOC, where the processed data are available to the GBM instrument team. The general scientific community may extract the Level 1 data from archives at the FSSC (http://fermi.gsfc.nasa.gov/ssc/data/).

GBM Level 1 processing primarily re-formats and reorganizes the Level-0 data. The gains of each detector are calibrated by monitoring the pulse-height channels of one or more background spectral lines. These gains are then be used to convert the raw detector pulse-height channels to an apparent energy. The Level 1 data consist of continuous and burst data. Continuous data are the rates in all GBM detectors in different energy bands, regardless of whether a burst has been detected. Burst data are the counts, rates, catalog information (e.g., fluence, duration, peak flux), and ancillary data necessary for analyzing the gamma-ray burst. The following three categories of data products are delivered to the *Fermi* Science Support Center: daily, burst and updates.

The daily data products (Table 4) consist of data that are produced continuously regardless of whether a trigger occurred. Thus these products are the count rates from all detectors, the monitoring of the detector calibrations (e.g., the position of the 511 keV line), and the spacecraft position and orientation. The underlying Level 0 data arrive with each Ku band telemetry downlink. However, the GIOC produces files of the resulting Level 1 data covering an entire calendar day (UT); these daily files are then sent to the FSSC in Flexible Image Transport System (FITS) format (http://fits.gsfc.nasa.gov/fits_intro.html). Consequently, the data latency is about one day: the first bit from the beginning of a calendar day may arrive a few hours after the day began while the last bit will be processed and added to the data product file a few hours after the day ended. These data products may be sent to the FSSC as they are produced, not necessarily in one package for a given day.

| GIOC Daily Data Products | | | | |
|---|---|---|---|---|
| ICD ID | Product | Description | Number of Files per Day | Size (bytes) |
| GS-001 | CTIME (daily version) | The counts accumulated every 0.256 s in 8 energy channels for each of the 14 detectors. | 14 | 230 MB (16 MB /file) |
| GS-002 | CSPEC (daily version) | The counts accumulated every 4.096 s in 128 energy channels for each of the 14 detectors. | 14 | 290 MB (20.6MB /file) |
| GS-005 | GBM gain and energy resolution history | History of the detector gains and energy resolutions; required for calculating DRMs. | 14 | 42kB (3kB/file) |
| GS-006 | position and attitude history | History of spacecraft position and attitude, required for calculating DRMs | 1 | 3MB |

Table 4. Daily Data products

The burst data products (Table 5) are the files pertaining to a given burst trigger that are produced and sent to the FSSC within a day after the burst. These include lists of counts, binned counts, and the DRMs and background spectra necessary to analyze the burst data. The burst products also include catalog files with summary data resulting from pipeline processing and a file with the TRIGDAT messages sent down over TDRSS immediately after a burst.

| | | **GIOC Burst Data Products** | | |
|---|---|---|---|---|
| ICD ID | Product | Description | Number of Files per Burst | Size (bytes) |
| GS-101 | CTIME (burst version) | For each detector, the counts accumulated every 0.064 s in 8 energy channels | 14 | 16MB (1.15 MB /file) |
| GS-102 | CSPEC (burst version) | For each detector, the counts accumulated every 1.024 s in 128 energy channels | 14 | 16MB (1.15 MB /file) |
| GS-103 | GBM TTE | Event data for the burst | 14 | 40-60MB (3-4.5 MB /file |
| GS-104 | GBM DRMs | 8 and 128 energy channel DRMs for all 14 detectors | 28 | 6 MB (0.4 MB /file) |
| GS-105 (non-burst trigger) | GBM Trigger Catalog Entry | Classification of GBM trigger with some characteristics | 1 | 20 kB |
| GS-106 (burst trigger) | GBM Burst or Spectral Catalog Entry | Values of the quantities describing the burst (e.g., durations, fluences) | 1 | 100-200 kB |
| GS-107 | GBM TRIGDAT | All the GBM's messages downlinked through TDRSS | 1 | 50-100 kB |
| GS-108 | GBM Background Files | Backgrounds for spectral fitting | 28 | 28kB (1kB /file) |

Table 5. Burst Data products

The final category of GIOC data products are those that are updated and sent to the FSSC periodically as required by new analysis (Table 6). These include calibrations that either do not change with time or change slowly. The catalogs—trigger, burst and spectral— are in this category. A preliminary version of the burst catalog file is distributed with the other burst data, while a number of updates will be provided subsequently as the data are reanalyzed, often with human intervention.

| \multicolumn{6}{c}{**GIOC Data Products Delivered as Updates**} |
| ICD ID | Product | Description | Number of Files | Frequency | Size (bytes) |
| --- | --- | --- | --- | --- | --- |
| GS-007 | GBM PHA Look-Up Tables | Tables of the correspondence between CTIME and CSPEC energy channels and the photopeak energy for each detector | 4 | Every ~6 months | 4kB (1kB/file) |
| GS-008 | GBM Calibration | Tables of fiducial detector response parameters from which the burst-specific DRMs are calculated | > 1000 | Every ~6 months | 100GB |
| GS-105 (non-burst trigger) | GBM Trigger Catalog Entry | Classification of GBM trigger with some characteristics | 1 | Updated periodically after initial file | 20 kB |
| GS-106 (burst trigger) | GBM Burst or Spectral Catalog Entry | Values of the quantities describing the burst (e.g., durations, fluences) | 1 | Updated periodically | 100-200 kB |

Table 6. Update Data products

### 6.4. *Burst Alert Processor*

The Burst Alert Processor (BAP) processes the TRIGDAT data in near-real time, and transmits burst locations and other parameters to users worldwide via the GRB Coordinates Network (GCN, http://gcn.gsfc.nasa.gov/) The primary BAP is located at the Mission Operations Center at GSFC, with an identical system available at the NSSTC in Huntsville for redundancy. The system is available at all times, with automatic switchover to the redundant system if the primary system is down. The BAP computes burst locations with better accuracy than is done on-board, since the ground system has the resources to include atmospheric and spacecraft scattering more accurately, uses a finer angular grid, and properly accounts for differences in burst spectra. Table 7 lists the types of GCN notices generated by the BAP.

| Notice Type | Information Provided |
|---|---|
| Fermi-GBM Alert | Trigger time, energy range, significance; detectors triggered, spacecraft lat. and long. |
| Fermi-GBM Flight Position | On-board computed localization (RA & Dec. plus spacecraft coordinates), classification, intensity, hardness ratio |
| Fermi-GBM Ground Position | On-ground computed localization (RA & Dec. plus spacecraft coordinates) |
| Fermi-GBM Ground Internal | Diagnostic data (sent to GBM team only) |

Table 7. GCN notices generated by the GBM Burst Alert Processor.

# 7. SUMMARY

The GBM detectors have been calibrated from 10 keV to 17.5 MeV using various gamma sources, and have been simulated over the entire energy range (8 keV to 40 MeV) using GEANT. The instrument has been operating successfully in orbit and triggering on gamma-ray bursts since July 11, 2008. The flight software maintains gain stability using background lines, classifies triggers and localizes gamma-ray bursts. The GBM Instrument Operations Center autonomously generates GCN alerts and provides GBM data to the *Fermi* Science Support Center. All level 1 and higher data from GBM are publicly available. The *Fermi* mission has a lifetime requirement of 5 years and a goal of 10 years.

Table 8 lists some of the key performance capabilities of GBM.

| | |
|---|---|
| Energy Resolution (FWHM) | ~15% at 100 keV; ~10% at 1MeV |
| High Rate Performance | < 2 % distortion at 50,000 cps |
| Energy Range | 8 keV to 40 MeV |
| On-board Trigger Threshold | 0.74 photons-cm$^{-2}$s$^{-1}$ |
| On-board Burst Location Error | < 15 degrees |
| Burst Trigger Rate | ~260 per year |
| Dead time per event | 2.6 microseconds |

Table 8. Key GBM performance capabilities.

ACKNOWLEDGEMENTS

The GBM science team gratefully acknowledges the outstanding engineering support from our institutions and contractors. The detectors were developed by Jena-Optronik.

The Power Supply Box was provided by Astrium. The Data Processing Unit was developed by Southwest Research Institute. Support for the German contribution was provided by the Bundesministerium für Bildung und Forschung (BMBF) via the Deutsches Zentrum für Luft- und Raumfahrt (DLR) under contract number 50 QV 0301. RMK & ASH acknowledge the support of the U.S. Department of Energy, National Nuclear Security Administration. A.J.v.d.H. was supported by an appointment to the NASA Postdoctoral Program at the MSFC, administered by Oak Ridge Associated Universities through a contract with NASA. The successful development of GBM is in large part due to the skill and dedication of our Lead Systems Engineer during the development phase, Mr. Fred Berry (1955-2006).

REFERENCES

Abdo, A. A. et al. 2009, Science, 323, 1688

Agostinelli, S. et al. 2003, NIM A, **506**, 250-303

Atwood, W. et al. 2009, ApJ, **697,** 1071

Band, D. et al., 1992, Experimental Astronomy, **2**, 307-330.

Band, D. et al. 1993, ApJ, **413**, 281

Briggs, M. S. et al. 2009, in "Gamma Ray Bursts, 6$^{th}$ Huntsville Symposium", AIP Conference Proceedings, **1133,** C. Meegan, N. Gehrels, C. Kouveliotou, eds., American Institute of Physics, Melville, NY.

Bildsten, L., et al. 1997, ApJS, **113**, 367

Bissaldi et al (2009) Exp. Astron., 24, 47

Cohen, E. & Piran, T. 1997, ApJ, **488**, L7

Fishman, G. J., & Austin, R. W. 1977, Nuclear Instruments and Methods, **140**, 193

Gonzalez, M. M., Dingus, B. L., Kaneko, Y., Preece, R. D., Dermer, C. D., & Briggs, M. S. 2003, Nature, **424**, 749

Harmon, B. A., et al. 2002, ApJS, **138**, 149

Hoover et al. 2009, in preparation

Hurley, K., et al. 1994, Nature, **372**, 652

Kaneko, Y. 2006 ApJS, **166**, 298

Kouveliotou, C., Norris, J. P., Wood, K. S., Cline, T. L., Dennis, B. R., Desai, U. D., & Orwig, L. E. 1992, ApJ, **392**, 179

Kouveliotou, C. et al. 1993 ApJ, **413**, L101

Leo, W.R. 1992, in "Techniques for Nuclear and Particle Physics Experiments", pub. Springer Verlag, p. 168.

Lichti, G. et al 2007, in "Gamma-Ray Bursts: Prospects for GLAST: Stockholm Symposium on GRB's", AIP Conference Proceedings, **906**, pp. 119-126, Axelsson, M. & Ryde, F., eds.

Meegan, C. A. et al. 2007, in "Proceedings of the First GLAST Symposium", AIP Press, AIP Conference Proceedings, **921**. pp 13-15, Ritz, S., Michelson, P., & Meegan, C, eds.

Meegan, C. A. et al. 2008, in "Gamma-Ray Bursts 2007; Proceedings of the Sante Fe Symposium", AIP Press, AIP Conference Proceedings, **1000**, pp 565-568, Galassi, M., Palmer, D., & Fenimore, E., eds.

Olivares, F. et al 2009, GCN Circular 9215

Pendleton, G.N., et al. 1995, NIM A, **364**, 567

Rosswog, S., Ramirez-Ruiz, E. & Davies, M. B. 2003, MNRAS, **345**, 1077

Tanvir, N. et al. 2009, GCN Circular 9219

Theis, L.M., Persyn, S.C., Johnson, M.A., Smith, K.D., Walls, B.J. and Epperly, M.E., 2006, IEEE Aerospace Conference, Paper #1217

Thoene, C. et al. 2009, GCN Circular 9216

Tinney, C., Stathakis, R., Cannon, R. & Galama, T. J. *IAU Circ. No. 6896* (1998).

Tsutsui, R., Nakamura, T., Yonetoku, D., Murakami, T., Tanabe, S., Kodama, Y. & Takahashi, K. 2009, MNRAS **394**, L31

Vedrenne, G. & Atteia, J-L 2009, Gamma-Ray Bursts (Chichester, UK, Praxis Publishing Ltd.)

Wijers R. A. M. J., Bloom, J. S., Bagla, J. S. & Natarajan, P. 1998, MNRAS **294**, L13

Woosley, S. E. 1993, ApJ, **405**, 273.